\documentclass[12pt]{article}

\usepackage[english]{babel}
\usepackage{subfigure}
\usepackage{latexsym}
\usepackage{graphicx}
\usepackage{epsfig}

\begin{document}

\title{Exact norm-conserving stochastic time-dependent Hartree-Fock}

\author{Luca Tessieri$^{1}$, Joshua Wilkie$^{2}$,
and Murat \c{C}etinba\c{s}$^{2}$ \\
$^{1}$ {\it \small Instituto de F\'{\i}sica y Matem\'{a}ticas} \\
{\it \small Universidad Michoacana de San Nicol\'{a}s de Hidalgo} \\
{\it \small 58060, Morelia, Michoac\'{a}n, Mexico} \\
$^{2}$ {\it \small Department of Chemistry,
Simon Fraser University} \\
{\it \small Burnaby, British Columbia V5A 1S6, Canada}}

\date{18th May 2004}
\maketitle

\begin{abstract}
We derive an exact single-body decomposition of the time-dependent
Schr\"{o}dinger equation for $N$ pairwise-interacting fermions. Each
fermion obeys a stochastic time-dependent norm-preserving wave equation.
As a first test of the method we calculate the low energy spectrum of
Helium. An extension of the method to bosons is outlined.
\end{abstract}

{Pacs 03.65.-w, 02.50.-r, 02.70.-c}

\section{Introduction}

Solution of the Schr\"{o}dinger equation for pairwise interacting identical 
fermions is a difficult computational problem with applications in many
areas of chemistry and physics. The development of accurate and
computationally efficient schemes for calculating the ground and excited
electronic states of molecules is a longstanding goal of theoretical
chemistry~\cite{Rev}. Electron dynamics plays an important role in molecular
electronics~\cite{Ratner} and atomic and molecular dynamics in strong
time-varying external fields~\cite{Rev}. The $N$-body problem for fermions
also arises in shell models in nuclear physics~\cite{Koonin}. Exact
strategies for $N$-body problems generally have computational costs which
scale exponentially with the number of particles. Here we show that exact
solutions of the $N$-fermion time-dependent Schr\"{o}dinger equation can be
obtained via a multi-configuration Hartree-Fock Ansatz in which the
single-particle wavefunctions for each configuration obey norm-conserving
stochastic wave equations.
Since all properties of the $N$-fermion problem can be calculated from the
exact time-evolving wavefunction, and since the computational costs
appear to scale favorably with the number of electrons, this method could
provide a useful alternative to other computational strategies such as
time-dependent density-functional theory~\cite{Rev} and auxiliary-field
quantum Monte Carlo~\cite{Koonin}. 

The technique of decomposing high dimensional deterministic equations into
lower dimensional stochastic wave equations was pioneered by Gisin and
Percival~\cite{GP} who were able to reduce deterministic master equations
for the density matrix into stochastic equations for a wavefunction. More
recently the same approach was used to reduce the $N$-boson Liouville
equation into one-boson stochastic wave equations~\cite{CCD}.
Similar decompositions have been obtained for fermions~\cite{JC} and
vibrations~\cite{Wilk}. Unfortunately, the norms of the single particle
stochastic wavefunctions grow exponentially for the boson and fermion
decompositions~\cite{CCD,JC}. This is the wave equation analog of the
``sign problem'' which plagues path integral Monte-Carlo
approaches~\cite{Koonin,QMC}. The decomposition for vibrations was derived
using a stochastic generalization of the time-dependent McLachlan
variational principle~\cite{Wilk}, and as a consequence the equations
conserve norm. Here we derive a similar norm conserving decomposition for
fermions. We demonstrate the use of the method by computing the low energy
spectrum of Helium. Finally, we explain how the same approach can be applied 
to bosons.

Before outlining the derivation in section~\ref{derswf} we summarise the
method here for readers who may not be interested in details.
In section~\ref{exact} we explicitly prove that the method is exact and that
the single body wave equations are norm conserving. Section~\ref{hel}
discusses an application of the method to Helium. In section~\ref{bos} we
explain how the method can be adapted for identical bosons.

We consider the general $N$ {\em identical} particle time-independent
Hamiltonian
\begin{equation}
{\cal H}_{N} = \sum_{i=1}^{N} H(i) + \sum_{i=1}^{N-1} \sum_{j=i+1}^{N} V(i,j)
\label{HAM}
\end{equation}
where $H(i)$ denotes the single body Hamiltonian instantiated for particle
$i$. For electrons in molecules $H=-\hbar^{2} \nabla^{2}/2m_{e} -
\sum_{k=1}^{M} Z_{k} e^{2}/|{\bf r}-{\bf R}_{k}|$, for example, where the
sum is over the nuclei of the molecule. The pairwise interaction $V(i,j)$
between particles $i$ and $j$ is represented via
\begin{equation}
V(i,j)=\sum_{s=1}^{p} \hbar\omega_{s} O_{s}(i)O_{s}(j)
\label{2-body}
\end{equation}
as a sum of products of dimensionless one-body (Hermitian or anti-Hermitian)
operators $O_{s}$. In section~\ref{dec} we prove that such an expansion is
always possible. The coefficients $\hbar\omega_{s}$ have units of energy and
may be positive or negative. This expansion is developed for the Coulomb
interaction $e^{2}/|{\bf r}_{i}-{\bf r}_{j}|$ in section~\ref{hel} (see also
Appendix A).
Extension of the method outined here to time-dependent Hamiltonians is
straightforward: simply replace $H$ and $O_{s}$ by their time-dependent
analogues in~(\ref{EQ2}) below.

A general initial $N$-fermion wavefunction can be written as a weighted sum
of Slater determinants of $N$ single particle wavefunctions. For our purposes
the single particle wavefunctions for a given determinant should be chosen so
that they are linearly independent and normalised but non-orthogonal.
Each Slater determinant can then be evolved independently. For simplicity we
now confine our attention
to one such initial state
\begin{equation}
|\Psi(0) \rangle = \beta A |\phi_{1}(0)\rangle |\phi_{2}(0)\rangle
\ldots |\phi_{N}(0)\rangle
\label{psi0}
\end{equation}
where $A$ is the anti-symmetrisation operator~\cite{Mess} and $\beta$ is a
normalisation constant. Here the position of a ``ket'' in the product
indicates which electron it refers to, i.e., for
$|\phi_{1}\rangle |\phi_{2}\rangle$ electron 1 is in state $|\phi_{1}\rangle$
and electron 2 is in state $|\phi_{2}\rangle$ while for
$|\phi_{2}\rangle |\phi_{1}\rangle$ electron 1 is in state $|\phi_{2}\rangle$
and electron 2 is in state $|\phi_{1}\rangle$. This convention allows us to
express some equations more simply than would otherwise be possible.

In our method the exact state $|\Psi(t)\rangle$ evolved from (\ref{psi0})
is constructed from the solutions $|\phi_{j}(t)\rangle$ of time-dependent
stochastic wave equations. Specifically, the exact $N$ fermion wavefunction
is expressed in terms of an average $M[\ldots ]$ via
\begin{equation}
|\Psi(t)\rangle = \beta M \left[ A|\phi_{1}(t)\rangle |\phi_{2}(t)\rangle
\ldots |\phi_{N}(t)\rangle \right]
\label{EQ1}
\end{equation}
where the $|\phi_{j}(t)\rangle$ obey It\^{o}-type~\cite{Hase} stochastic
equations
\begin{equation}
\begin{array}{ccl}
d|\phi_{j}\rangle & = & \displaystyle 
\left( - \frac{i}{\hbar} H |\phi_{j}\rangle +
\frac{i}{2} \sum_{k \neq j}\sum_{s=1}^{p} \omega_{s}
\langle O_{s}\rangle_{j} \langle O_{s}\rangle_{k} |\phi_{j} \rangle \right. \\
& - & \displaystyle \left. i \sum_{k \neq j} \sum_{s=1}^{p}
\omega_{s} \langle O_{s}\rangle_{k} O_{s} |\phi_{j} \rangle \right) dt +
\sum_{s=1}^{p} \sqrt{-i \omega_{s}} \left( O_{s} - \langle O_{s}\rangle_{j}
\right) |\phi_{j} \rangle dW_{s} \\
& - & \displaystyle
\sum_{k \neq j} \sum_{s=1}^{p} |\omega_{s}|\langle \phi_{j}|\phi_{j} \rangle
\frac{\langle O_{s}^{\dag}O_{s}\rangle_{j} - |\langle O_{s}\rangle_{j}|^{2}}
{2(N-1) {\rm Re} \left\{ \langle \phi_{j}|\phi_{k} \rangle \right\} }
|\phi_{k} \rangle dt
\end{array}
\label{EQ2}
\end{equation}
for $j=1,\dots, N$. Here we use a notation where $\langle F\rangle_{j} =
\langle \phi_{j}|F|\phi_{j} \rangle/\langle \phi_{j}|\phi_{j} \rangle$ for
any single-body operator $F$.
For notational simplicity the explicit time dependence of $|\phi_{j}\rangle$
and the stochastic random variables $dW_{s}$ has not been indicated. [Note
that in the case of electrons $|\phi_{j}\rangle$ are similar to the spin-orbit
single particle wavefunctions of Hartree-Fock.]
The symbols $dW_{s}(t)$ represent independent normally distributed real
stochastic differentials with
\begin{eqnarray}
M \left[ d W_{s} (t)\right] = 0 & \mbox{and} &
M \left[ d W_{r}(t) d W_{s} (t)\right] = \delta_{rs} \; dt .
\label{RVAR}
\end{eqnarray}
The second condition imposes statistical independence of the stochastic
differentials. 

Imagine a sequence of time steps all of equal length $dt$ such that $t=mdt$
for some integer $m$. At each time step a set of stochastic differentials is
sampled from the normal distribution 
\begin{displaymath}
P\left(d {\bf W}(l dt)\right) = [1/(2\pi dt)]^{p/2}
\exp \{ -d{\bf W}(ldt)\cdot d{\bf W}(ldt)/2dt \}
\end{displaymath}
where $d{\bf W}(ldt)=(dW_{1}(ldt),\dots, dW_{p}(ldt))$ is the vector of
stochastic differentials ($p$ is the number of components of $d{\bf W}$).
Note that $l$ runs from 1 to $m$.
The expectation~(\ref{EQ1}) at any time $t$ can thus be represented in the
form
\begin{displaymath}
|\Psi(t)\rangle =\beta \prod_{l=1}^m \int d^{p}W(ldt)
P\left(d {\bf W}(l dt)\right)
A|\phi_{1}(t)\rangle |\phi_{2}(t)\rangle
\ldots |\phi_{N}(t)\rangle
\end{displaymath}
and Monte-Carlo sampling of the integrals then yields the stochastic paths
generated by Eqs.~(\ref{EQ2}). Each time sequence of sampled stochastic
differentials defines one set of stochastic variables $W_{s}(t)$ (i.e.,
Wiener process). Each realisation of the set of stochastic variables
$W_{s}(t)$ as a function of time thus yields one Slater determinant in the
average $M[\ldots]$.
Since single particle norms are conserved each Slater determinant is equally
weighted in the average, and error in the mean will scale as $1/\sqrt{L}$
where $L$ is the number of realisations.

The single particle wavefunctions on the right hand side of (\ref{EQ2}) are
independent of the stochastic differentials $dW_{s}(t)$ and so averages such
as $M[F(\phi_1(t),\dots,\phi_N(t))g(dW_{1}(t),\dots, dW_{p}(t))]$ can be
calculated via the simplified formula $M[F(\phi_1(t),\dots,\phi_N(t))]
M[g(dW_{1}(t),\dots, dW_{p}(t))]$. This fact is implicit in proofs of
norm-conservation and exactness outlined in section~\ref{exact}.

The fact that all matrix elements (e.g. $\langle \phi_{j}|O_{s}|\phi_{j}
\rangle $) in the stochastic equations involve single particle operators,
and the sum over index $k \neq j$ for each $|\phi_{j} \rangle$, show that
the computational costs will scale at least quadratically with the number of
electrons. For implementations similar to that for He, discussed in
section~\ref{hel}, the number of terms in the two-body expansion {\em in
principle} scales as the square of the number of electrons (in practice
many $\hbar\omega_s$ may be small or zero which could improve the scaling
of the method), and hence evaluation of all $\langle \phi_{j}|O_{s}|\phi_{j}
\rangle$ for each $j$ requires $N^{4}$ operations, making the method scale
as $O(N^{5})$ overall. 
The precise scaling is obviously model dependent but computational costs
should be somewhere in the range $O(N^{2})$ to $O(N^{5})$. Most alternative
exact approaches have computational costs which scale exponentially with the
number of electrons. 

The most important properties of the stochastic decomposition~(\ref{EQ1})
and~(\ref{EQ2}) are its exactness and its norm conservation. 

Equations (\ref{EQ2}) conserve norm in the mean (i.e.,
$M[\langle \phi_{j}(t)|\phi_{j}(t)\rangle]=1$) which gives our decomposition
distinct numerical advantages over other decompositions in which the mean
norm grows exponentially~\cite{JC}. In addition, our method conserves norm
exactly for each individual stochastic realisation (see section~\ref{exact}).
Note that the norm of $A|\phi_{1}(t)\rangle |\phi_{2}(t)\rangle \ldots
|\phi_{N}(t)\rangle $ is {\em not} conserved by our method because the single
particle states are non-orthogonal. This however presents no problem
numerically.

Using the It\^{o} calculus~\cite{Hase} we also show in section~\ref{hel} that
\begin{equation}
\begin{array}{ccl}
d|\Psi(t)\rangle & = & \displaystyle
\beta M \left[ \sum_{j=1}^{N} A |\phi_{1}(t)\rangle \ldots
|d\phi_{j}(t)\rangle \ldots |\phi_{N}(t)\rangle \right. \\
& + & \displaystyle \left. \sum_{j=1}^{N-1} \sum_{k=j+1}^{N} 
A |\phi_{1}(t)\rangle \ldots |d\phi_{j}(t)\rangle \ldots |d\phi_{k}(t)\rangle
\ldots |\phi_{N}(t) \rangle \right] \\
& = & \displaystyle -\frac{i}{\hbar}{\cal H}_{N}|\Psi(t)\rangle ~dt
\end{array}
\label{tchange}
\end{equation}
which implies that the method is exact for all forms of our equations.

Explicit time-dependence of the $N$-fermion wavefunction is of direct
interest in many chemical problems. Energies can be extracted via the
Fourier transform of the time auto-correlation function
$\langle \Psi(0)|\Psi(t)\rangle$.
In practice, one computes the function
\begin{equation}
I(E) = \frac{1}{\pi\hbar} {\rm Re} \int_{0}^{T} \langle \Psi(0)|\Psi(t)\rangle
\exp \left(\frac{iEt}{\hbar} \right) \; dt
\simeq \langle \Psi(0)|\delta(E-{\cal H}_{N})|\Psi(0)\rangle
\label{spectrum}
\end{equation}
which will have maxima at the true energies when the end point of integration
$T$ is sufficiently large.
The method therefore also provides access to spectral information and in fact
it is straightforward to generalize~(\ref{spectrum}) so that states of
specific parity can be extracted. Eigenfunctions can also be obtained.

\section{Single body decomposition of pairwise interaction}
\label{dec}

Consider a general two-body interaction $V(1,2)$. We will now show that it
can be expanded in products of one-body interactions according to
Eq.~(\ref{2-body}). Let $|i\rangle$ with $i=1,2,\dots$ denote a complete
basis of the one-body space. Then $|i_{1};i_{2}\rangle =
|i_{1}\rangle |i_{2}\rangle$ for $i_{1}, i_{2} = 1,2,\dots$ will be a complete
basis of the two-body space. Here again we employ the convention that the
position of a ``ket'' in a product indentifies the fermion. It follows then
that we may represent the interaction via
\begin{equation}
V(1,2) = \sum_{i_{1}=1}^{\infty} \sum_{i_{2}=1}^{\infty}
\sum_{j_{1}=1}^{\infty} \sum_{j_{2}=1}^{\infty} |i_{1};i_{2}\rangle
\langle i_{1};i_{2}|V(1,2)|j_{1};j_{2}\rangle \langle j_{1};j_{2}|
\label{expansion}
\end{equation}
where we have inserted closure relations for the two-body space on either side.

Define a bijective application $\sigma : {\bf N}^{2} \rightarrow {\bf N}$
which maps each couple of integers $(i,j)$ in a unique integer $\sigma (i,j)$.
Then we can introduce new composite indices $\sigma_{1}= \sigma (i_{1},j_{1})$
for body 1 and $\sigma_{2} = \sigma(i_{2},j_{2})$ for body 2 with $\sigma_{1}$
and $\sigma_{2}$ taking integer values $1,2,\dots$.
We may then define matrix elements
\begin{displaymath}
{\cal V}_{\sigma_{1},\sigma_{2}} =
\langle i_{1};i_{2}|V(1,2)|j_{1};j_{2}\rangle
\end{displaymath}
which are symmetric under the interchange of $\sigma_{1}$ and $\sigma_{2}$.
This symmetry reflects the indistinguishability of the particles.
Diagonalising ${\cal V}$ then gives
\begin{equation}
{\cal V}_{\sigma_{1},\sigma_{2}} = \sum_{s=1}^{\infty} \hbar \omega_{s}
Q_{\sigma_{1},s} Q_{\sigma_{2},s}
\label{2body}
\end{equation}
where $\hbar\omega_{s}$ are the eigenvalues and $Q_{\sigma,s}$ are the
dimensionless matrix elements of the orthogonal transformation. With a
slight change of notation and using the inverse of the mapping $\sigma_{1}
= \sigma(i_{1},j_{1})$ we may then write
\begin{equation}
Q_{\sigma_{1},s}=\langle i_{1}|O_{s}|j_{1}\rangle
\label{one-body}
\end{equation}
which defines the one body operator $O_{s}$. Since $V(1,2)$ is Hermitian it
follows that each $O_{s}$ must be either Hermitian or anti-Hermitian.
The eigenvalues $\hbar\omega_{s}$ may be positive or negative.

Substituting~(\ref{one-body}) into~(\ref{2body}), and~(\ref{2body})
into~(\ref{expansion}) gives
\begin{eqnarray*}
V(1,2) & = & \sum_{i_{1}=1}^{\infty} \sum_{i_{2}=1}^{\infty}
\sum_{j_{1}=1}^{\infty} \sum_{j_{2}=1}^{\infty}
|i_{1};i_{2}\rangle \sum_{s=1}^{\infty} \hbar\omega_{s}
\langle i_{1}|O_{s}|j_{1}\rangle \langle i_{2}|O_{s}|j_{2}\rangle
\langle j_{1};j_{2}| \\
& = & \sum_{s=1}^{\infty} \hbar\omega_{s} \sum_{i_{1}=1}^{\infty}
\sum_{i_{2}=1}^{\infty} \sum_{j_{1}=1}^{\infty} \sum_{j_{2}=1}^{\infty}
|i_{1};i_{2}\rangle  \langle i_{1};i_{2}|O_{s}(1)O_{s}(2)|j_{1};j_{2}\rangle
\langle j_{1};j_{2}|.
\end{eqnarray*}
Finally removing the closure relations gives
\begin{displaymath}
V(1,2) = \sum_{s=1}^{\infty} \hbar\omega_{s} O_{s}(1) O_{s}(2)
\end{displaymath}
which is the desired expansion. 

In practice a finite basis set is more practical than a complete one but the
same considerations apply except that the sum will terminate at some finite
value $p$.

\section{Derivation of stochastic wave equations}
\label{derswf}

We originally derived the stochastic decomposition discussed above using
the stochastic McLachlan variational principle developed in Ref.~\cite{Wilk}.
Here we present a more direct argument. For simplicity we initially focus
on just two fermions with the simplest possible interaction. Consider then
the restricted two fermion Hamiltonian
\begin{equation}
{\cal H}_{2} = H(1) + H(2) + \hbar\omega \, O(1)O(2)
\label{simpleH}
\end{equation}
and a normalised initial wavefunction of the form 
\begin{displaymath}
|\Psi(0)\rangle =
\beta \left( |\phi_{1}(0)\rangle |\phi_{2}(0)\rangle
- |\phi_{2}(0)\rangle |\phi_{1}(0)\rangle \right)
\end{displaymath}
where $\beta=1/\sqrt{2(1-|\langle \phi_{1}(0)|\phi_{2}(0)\rangle|^{2})}$
is a normalisation factor, and $|\phi_{1}(0)\rangle$ and
$|\phi_{2}(0)\rangle$ are normalised but non-orthogonal states, i.e.,
\begin{eqnarray*}
\langle \phi_{1}(0)|\phi_{1}(0) \rangle =
\langle \phi_{2}(0)|\phi_{2}(0)\rangle = 1 & \mbox{and} &
\langle \phi_{1}(0)|\phi_{2}(0)\rangle \neq 0.
\end{eqnarray*}
Note the antisymmetric form of the initial wavefunction.

We wish to find stochastic equations for $|\phi_{1}(t)\rangle$ and
$|\phi_{2}(t)\rangle$ such that the exact solution $| \Psi(t) \rangle$ of
the Schr\"{o}dinger equation
\begin{displaymath}
d | \Psi(t) \rangle = - \frac{i}{\hbar} {\cal H}_{2} |\Psi(t) \rangle dt
\end{displaymath}
can be written as the expectation value
\begin{equation}
|\Psi(t)\rangle = \beta M \left[ |\Phi(t) \rangle \right]
\label{soln}
\end{equation}
of the antisymmetric stochastic vector
\begin{equation}
|\Phi(t) \rangle =
|\phi_{1}(t)\rangle |\phi_{2}(t) \rangle -
|\phi_{2}(t)\rangle |\phi_{1}(t)\rangle.
\label{defn}
\end{equation}
We will also require that the stochastic wave equations conserve norm
\begin{displaymath}
\langle \phi_{1}(t)|\phi_{1}(t)\rangle =
\langle \phi_{2}(t)|\phi_{2}(t)\rangle = 1.
\end{displaymath}

To achieve norm conservation the single fermion wavefunctions must satisfy
the condition
\begin{equation}
d\left( \langle \phi_{i} | \phi_{i} \rangle \right) =
\langle d\phi_{i}|\phi_{i}\rangle + \langle \phi_{i}|d\phi_{i}\rangle
+ \langle d\phi_{i}|d\phi_{i}\rangle = 0
\label{cond1}
\end{equation}
for $i=1,2$. Since $d|\phi_{i}(t)\rangle$ will have a term proportional to
a change $dW(t)$ in a stochastic process $W(t)$ with
$M[dW(t)^{2}] = dt$ and $M[dW(t)] = 0$ , there will naturally be
terms proportional to $dW(t)^{2}$ (which is of order $dt$) in
condition~(\ref{cond1}). Hence it may prove useful to have a term
proportional to $dW(t)^{2}$ in $d|\phi_{i}(t)\rangle$ in order to conserve
norm. Our wave equations should therefore be of the form
\begin{equation}
d|\phi_{i}\rangle = |v_{i}\rangle dt + |u_{i}\rangle dW +
|w_{i}\rangle dW^{2}
\label{seqns}
\end{equation}
where all quantities depend on the time $t$. With this form of the
stochastic differential $d|\phi_{i}\rangle$, condition~(\ref{cond1})
can be written as
\begin{equation}
2{\rm Re} \left\{ \langle \phi_{i}|v_{i} \rangle \right\} dt +
2{\rm Re} \left\{ \langle \phi_{i}|u_{i} \rangle \right\} dW +
\left( 2 {\rm Re} \left\{ \langle \phi_{i}|w_{i}\rangle \right\} +
\langle u_{i}|u_{i}\rangle \right) dW^{2} = 0
\label{cond2}
\end{equation}
and the coefficients of $dt$, $dW$ and $dW^{2}$ must independently
vanish.

In order to reproduce the interaction term of Hamiltonian~(\ref{simpleH})
we must have a term in $|u_{i}\rangle$ which is proportional to
$O|\phi_{i}\rangle$. To make the coefficient of $dW$ vanish in
Eq.~(\ref{cond2}) it would thus be sufficient to choose 
\begin{equation}
|u_{i}\rangle = 
\sqrt{-i\omega} \left( O - \langle O\rangle_i \right) |\phi_{i} \rangle
\label{term1}
\end{equation}
eliminating one of the unknowns in Eq.~(\ref{seqns}). Here
$\langle O\rangle_{i} = \langle \phi_{i}|O|\phi_{i}\rangle/
\langle \phi_{i}|\phi_{i}\rangle$ where we keep the factor of
$\langle \phi_{i}|\phi_{i}\rangle$ explicit even though it is unity.

To make the coefficient of $dW^{2}$ vanish in Eq.~(\ref{cond2}) we
can choose 
\begin{equation}
\begin{array}{ccl}
|w_{1}\rangle & = & \displaystyle
-|\omega|\frac{\langle \phi_{1}|\phi_{1}\rangle[\langle O^{\dag}O\rangle_1 -
|\langle O \rangle_1|^{2}]}
{2 {\rm Re} \left\{ \langle \phi_{1}|\phi_{2} \rangle \right\}}
|\phi_{2}\rangle \\
|w_{2}\rangle & = & \displaystyle
-|\omega|\frac{\langle \phi_{2}|\phi_{2} \rangle[\langle O^{\dag}O\rangle_2 -
|\langle O\rangle_2|^{2}]}
{2 {\rm Re} \left\{ \langle \phi_{1}|\phi_{2} \rangle \right\}}
|\phi_{1}\rangle
\end{array}
\label{term2}
\end{equation}
since the $|w_{i}\rangle$ terms were included precisely for this purpose.
Clearly $\phi_{1}$ and $\phi_{2}$ must be non-orthogonal initially and a
declining overlap will cause an increase of~(\ref{term2}) for each mode
thereby restoring the overlap.

Finally, we need to find $|v_{i}\rangle$. Clearly, there should be a term
like $-(i/\hbar) H|\phi_{i}\rangle$ to reproduce the single particle terms
of Hamiltonian~(\ref{simpleH}). There could also be a term like
$O|\phi_{i}\rangle$. So assume that $|v_{i}\rangle$ will take the form
\begin{equation}
|v_{i}\rangle = -(i/\hbar) H |\phi_{i}\rangle + a_{i} |\phi_{i} \rangle
+ b_{i} O |\phi_{i}\rangle
\label{term3}
\end{equation}
where $a_{i}$ and $b_{i}$ are unknowns. To make the coefficient of $dt$
vanish in Eq.~(\ref{cond2}) it is necessary that ${\rm Re}\{ a_{i} \}
+ {\rm Re}\{b_{i}\} \langle \phi_{i}| O |\phi_{i} \rangle = 0$. Hence we
can probably set the real parts of $a_{i}$ and $b_{i}$ to zero. To determine
their imaginary parts we consider the expectation of the differential of
the vector~(\ref{defn}) which, because of condition~(\ref{soln}),
must be equal to the differential of the vector $|\Psi(t)\rangle$, so
that one has
\begin{eqnarray}
d|\Psi(t)\rangle &= &
\beta M [ |d \phi_{1}(t)\rangle |\phi_{2}(t)\rangle +
|\phi_{1}(t) \rangle |d\phi_{2}(t)\rangle +
|d\phi_{1}(t) \rangle |d\phi_{2}(t)\rangle\nonumber \\ 
&-&|d\phi_{2}(t)\rangle |\phi_{1}(t)\rangle -
|\phi_{2}(t)\rangle |d\phi_{1}(t) \rangle -
|d\phi_{2}(t)\rangle |d\phi_{1}(t)\rangle ] .
\label{diff}
\end{eqnarray}
Replacing the differential terms $d|\phi_{i}\rangle$ in the right-hand
side of the previous equation with expression~(\ref{seqns}) and making
use of the results~(\ref{term1}) and~(\ref{term2}) as well as of
Ansatz~(\ref{term3}), after some algebra one obtains
\begin{equation}
\begin{array}{ccl}
d|\Psi(t)\rangle & = & \displaystyle
\beta M \left[ -(i/\hbar) (H(1) + H(2)) |\Phi(t) \rangle dt
- i\omega O(1) O(2) |\Phi(t) \rangle dW^{2} \right. \\
& + & \displaystyle
(a_{1} + a_{2}) |\Phi(t) \rangle dt -
i\omega \langle O\rangle_1
\langle O \rangle_2 |\Phi(t) \rangle dW^{2} \\
& + & \displaystyle
\sqrt{-i\omega} \left( O(1) + O(2) - \langle O \rangle_1
- \langle O\rangle_2 \right) |\Phi(t)\rangle dW \\
& + & \left(b_{1} dt + i\omega \langle O \rangle_2
dW^{2} \right)| O \phi_{1}(t) \rangle |\phi_{2}(t) \rangle \\
& + & \displaystyle
\left( b_{2} dt + i\omega \langle O\rangle_1
dW^{2} \right) |\phi_{1}(t)\rangle | O \phi_{2}(t)\rangle \\
& - & \displaystyle
\left( b_{1} dt + i\omega \langle O\rangle_2
dW^{2} \right) |\phi_{2}(t)\rangle | O \phi_{1}(t) \rangle \\
& - & \displaystyle \left.
\left( b_{2} dt + i\omega \langle O\rangle_1
dW^{2} \right) | O \phi_{2}(t)\rangle |\phi_{1}(t) \rangle \right].
\end{array}
\label{diff2}
\end{equation}
Using condition~(\ref{soln}) and the facts that $M[dW]=0$ and
$M[dW^{2}]=dt$, and assigning
\begin{eqnarray*}
a_{1} = a_{2} = \frac{i\omega}{2} \langle O \rangle_1
\langle O \rangle_2, ~~~~b_{1} = -i\omega \langle O \rangle_2 & \mbox{and} &
b_{2} = -i\omega \langle O \rangle_1 ,
\end{eqnarray*}
we then find that Eq.~(\ref{diff2}) reduces to
$d|\Psi(t)\rangle = - (i/\hbar) {\cal H}_2|\Psi(t)\rangle dt$
which is the exact Schr\"{o}dinger equation in differential form.
Hence we have found exact stochastic wave equations of the form
\begin{equation}
\begin{array}{ccl}
d|\phi_{1}\rangle & = & \displaystyle
\left( -\frac{i}{\hbar} H|\phi_{1} \rangle - i\omega\langle O \rangle_{2} O 
|\phi_{1} \rangle + \frac{i\omega}{2}) \langle O \rangle_{1}
\langle O \rangle_{2} |\phi_{1} \rangle \right) dt \\
& + & \displaystyle
\sqrt{-i\omega} \left( O - \langle O \rangle_1 \right)
|\phi_{1} \rangle dW - |\omega| \frac{\langle \phi_{1}|\phi_{1}\rangle
\left[\langle O^{\dag}O\rangle_{1} - |\langle O \rangle_1|^{2}\right]}
{2 {\rm Re} \left\{ \langle \phi_{1} |\phi_{2} \rangle \right\}}
|\phi_{2} \rangle dW^{2} \\
d|\phi_{2}\rangle & = & \displaystyle
\left( -\frac{i}{\hbar} H|\phi_{2} \rangle - i\omega\langle O \rangle_{1} O
|\phi_{2}\rangle + \frac{i\omega}{2} \langle O \rangle_{1}
\langle O\rangle_2 |\phi_{2}\rangle \right) dt \\
& + & \displaystyle
\sqrt{-i\omega} \left( O - \langle O\rangle_{2} \right)
|\phi_{2}\rangle dW - |\omega|\frac{\langle \phi_{2}|\phi_{2}\rangle
\left[\langle O^{\dag}O \rangle_{2} - |\langle O \rangle_2|^{2}\right]}
{2 {\rm Re} \left\{ \langle \phi_{1}|\phi_{2} \rangle \right\} }
|\phi_{1} \rangle dW^{2}
\end{array}
\label{Oeqns}
\end{equation}
which conserve norm by construction. Since terms of order $dW^{3}$ and
higher are of no importantance and since the average of $dW^{2}$ is $dt$,
it is possible to make this replacement in Eqs.~(\ref{Oeqns}) with no loss
of accuracy or generality~\cite{Hase} giving
\begin{equation}
\begin{array}{ccl}
d|\phi_{1}\rangle & = & \displaystyle
\left(-\frac{i}{\hbar} H|\phi_{1} \rangle - i\omega \langle O \rangle_{2} O
|\phi_{1} \rangle + \frac{i\omega}{2}) \langle O \rangle_{1}
\langle O \rangle_{2} |\phi_{1} \rangle \right) dt \\
& + & \displaystyle
\sqrt{-i\omega} \left( O - \langle O \rangle_{1} \right)
|\phi_{1} \rangle dW -|\omega|\frac{\langle \phi_{1}|\phi_{1}\rangle
\left[ \langle O^{\dag}O\rangle_{1} - |\langle O \rangle_{1}|^{2}\right]}
{2 {\rm Re} \left\{ \langle \phi_{1} |\phi_{2} \rangle \right\}}
|\phi_{2} \rangle dt \\
d|\phi_{2}\rangle & = & \displaystyle
\left( -\frac{i}{\hbar} H|\phi_{2} \rangle - i\omega \langle O \rangle_{1} O
|\phi_{2}\rangle + \frac{i\omega}{2} \langle O \rangle_{1}
\langle O \rangle_{2} |\phi_{2} \rangle \right) dt \\
& + & \displaystyle
\sqrt{-i\omega} \left( O - \langle O\rangle_{2} \right)
|\phi_{2}\rangle dW - |\omega|\frac{\langle \phi_{2}|\phi_{2}\rangle
\left[\langle O^{\dag}O\rangle_{2} - |\langle O \rangle_2|^{2}\right]}
{2 {\rm Re} \left\{ \langle \phi_{1}|\phi_{2} \rangle \right\} }
|\phi_{1} \rangle dt.
\end{array}
\label{Oeqns2}
\end{equation}

Generalisation of~(\ref{Oeqns2}) to the full pairwise interaction gives a
special case of~(\ref{EQ1}) and~(\ref{EQ2}). We thus proceed directly in the
next section to consideration of the $N$-fermion problem with full pairwise
interaction.

\section{Exactness and conservation of one-body norm}
\label{exact}

Consider conservation of norm first. To be norm conserving Eq.~(\ref{EQ2})
must satisfy the constraint
\begin{displaymath}
d\left(\langle \phi_{j}(t)|\phi_{j}(t)\rangle\right) =
\langle d\phi_{j}(t)|\phi_{j}(t)\rangle +
\langle \phi_{j}(t)|d\phi_{j}(t)\rangle +
\langle d\phi_{j}(t)|d\phi_{j}(t)\rangle =0
\end{displaymath}
for $j=1,\dots, N$ or equivalently that
\begin{eqnarray*}
dM[\langle \phi_{j}(t)|\phi_{j}(t)\rangle] = 0 & \mbox{and} &
dM[\langle \phi_{j}(t)|\phi_{j}(t)\rangle^2] = 0.
\end{eqnarray*}
Substituting~(\ref{EQ2}) in $dM[\langle \phi_{j}(t)|\phi_{j}(t)\rangle]$ gives
\begin{displaymath}
 M[\sum_{s=1}^{p} |\omega_{s}|\langle \phi_{j}|\phi_{j} \rangle
\left( \langle O_{s}^{\dag}O_{s}\rangle_{j} - |\langle O_{s}\rangle_{j}|^{2}
\right)(dW_{s}^{2}-dt)]
\end{displaymath}
which vanishes.
Similarly,
\begin{displaymath}
\begin{array}{ccl}
dM \left[\langle \phi_{j}(t)|\phi_{j}(t)\rangle^{2}\right] & = &
\displaystyle
M\left[2 \langle \phi_{j}(t)|\phi_{j}(t)\rangle
d \left(\langle \phi_{j}(t)|\phi_{j}(t)\rangle \right) \right. \\
& + & \displaystyle \left.
2 |\langle \phi_{j}(t)|d\phi_{j}(t)\rangle|^{2} +
\langle \phi_{j}(t)|d\phi_{j}(t)\rangle^{2} +
\langle d\phi_{j}(t)|\phi_{j}(t)\rangle^{2} \right]
\end{array}
\end{displaymath}
 which then gives
\begin{displaymath}
 M\left[2\sum_{s=1}^{p} |\omega_{s}|\langle \phi_{j}|\phi_{j} \rangle^{2}
\left( \langle O_{s}^{\dag}O_{s}\rangle_{j} - |\langle O_{s}\rangle_{j}|^{2}
\right) \left(dW_{s}^{2}-dt\right) + O(dt^{2}) \right]
\end{displaymath}
which vanishes as $dt\rightarrow 0$. Hence norm is exactly conserved for
individual stochastic realisations as well as in the mean.

Now consider the issue of exactness of the decomposition.
Substituting~(\ref{EQ2}) into Eq.~(\ref{tchange}) we see that the
term of~(\ref{EQ2}) proportional to $|\omega_{s}|$ makes no contribution
because the Slater determinants have two identical single particle
orbitals and hence vanish. The term of~(\ref{EQ2}) proportional to $dW_{s}$
makes no contribution to the first term of~(\ref{tchange}) because
$M[dW_{s}]=0$. The first three terms of~(\ref{EQ2}) contribute
\begin{equation}
\begin{array}{l}
\displaystyle
M \left[ -\frac{i}{\hbar} \sum_{j=1}^{N-1} A |\phi_{1}\rangle \ldots
|H \phi_{j}\rangle \ldots |\phi_{N} \rangle \right.\\
\displaystyle
+ i \sum_{j=1}^{N-1} \sum_{k=j+1}^{N} \sum_{s=1}^{p} \omega_{s}
\langle O_{s}\rangle_j \langle O_{s}\rangle_{k}
A |\phi_{1} \rangle \ldots |\phi_{N} \rangle \\
\displaystyle
-i \sum_{j=1}^{N-1} \sum_{k=j+1}^{N} \sum_{s=1}^p \omega_{s} \langle
O_{s} \rangle_j A |\phi_{1} \rangle \ldots |O_{s}
\phi_{k}\rangle \ldots |\phi_{N}\rangle \\
\displaystyle \left.
- i \sum_{j=1}^{N-1} \sum_{k=j+1}^{N} \sum_{s=1}^{p} \omega_{s} \langle
O_{s}\rangle_k A |\phi_{1} \rangle \ldots |O_{s}
\phi_{j} \rangle \ldots |\phi_{N}\rangle \right] dt
\end{array}
\label{contrib1}
\end{equation}
to the first term of~(\ref{tchange}). The non-vanishing contributions of
the second term in Eq.~(\ref{tchange}) are
\begin{equation}
\begin{array}{l}
\displaystyle
M \left[ -i \sum_{j=1}^{N-1} \sum_{k=j+1}^{N} \sum_{s=1}^{p} \omega_{s}
A |\phi_{1}\rangle \ldots |O_{s} \phi_{j}\rangle \ldots
|O_{s} \phi_{k}\rangle \ldots |\phi_{N} \rangle \right. \\
\displaystyle
- i \sum_{j=1}^{N-1} \sum_{k=j+1}^{N} \sum_{s=1}^{p} \omega_{s} \langle
O_{s}\rangle_{j} \langle O_{s}\rangle_{k}
A |\phi_{1} \rangle \ldots |\phi_{N} \rangle \\
\displaystyle
+ i \sum_{j=1}^{N-1} \sum_{k=j+1}^{N} \sum_{s=1}^p \omega_{s} \langle
O_{s}\rangle_{j} A |\phi_{1} \rangle \ldots
|O_{s} \phi_{k}\rangle \ldots |\phi_{N}\rangle \\
\displaystyle \left.
+ i \sum_{j=1}^{N-1} \sum_{k=j+1}^{N} \sum_{s=1}^{p} \omega_{s} \langle
O_{s}\rangle_k A |\phi_{1} \rangle \ldots |O_{s}
\phi_{j} \rangle \ldots |\phi_{N}\rangle \right] dt.
\label{contrib2}
\end{array}
\end{equation}
The first terms of~(\ref{contrib1}) and~(\ref{contrib2}) combine to give
$-(i/\hbar){\cal H}_N|\Psi(t)\rangle dt$. The second terms of~(\ref{contrib1})
and~(\ref{contrib2}) cancel as do the third and fourth terms of the
respective equations. These considerations thus show that
$d|\Psi(t)\rangle=-(i/\hbar){\cal H}_N|\Psi(t)\rangle dt$ and hence that the
method is exact.

\section{Application to ${\rm He}$}
\label{hel}

Here we apply Eqs.~(\ref{EQ1}) and~(\ref{EQ2}) to the problem of calculating
the low energy spectrum of Helium as a first test of the method. Clearly we
need a basis in which to represent the single electron wavefunctions. We also
need a decomposition of the form~(\ref{2-body}) for the Coulomb interaction. 

We choose to represent the single particle wavefunctions in a finite basis of
states $|n,l,m,\tau\rangle=|\psi_{n,l,m}\rangle \otimes |\tau\rangle$ where
$\psi_{n,l,m}$ are the exact orbital eigenfunctions of the ${\rm He}^+$ ion
and $|\tau\rangle=|\pm\rangle$ for $\tau=\pm$ are the spin-1/2 eigenstates.
Here $n=1,2,\dots$, $l=0,\dots n-1$, and $m=-l,\dots ,0,\dots l$ are the
allowed values of the quantum numbers. That is, the orbital parts of these
basis functions are exact eigenfunctions of the single-body Hamiltonian
$-\hbar^{2} \nabla^{2}/2m_{e}-2e^{2}/r$ with eigenvalues $E_{n}=-2/n^{2}$
in atomic units (i.e. $\hbar=1$, $m_{e}=1$, and $e=1$). The functional forms
in the coordinate representation are
\begin{equation}
\langle {\bf r}|\psi_{n,l,m}\rangle = R_{n,l}(r)Y_{l,m}(\theta,\phi)
\label{basis}
\end{equation}
where $Y_{l,m}(\theta,\phi)$ are the usual spherical harmonics (i.e.,
eigenstates of angular momentum) and the radial functions $R_{n,l}(r)$ are
\begin{displaymath}
R_{n,l}(r)= \frac{4}{n^{2}}\sqrt{2\frac{(n-l-1)!}{(n+l)!}}
e^{-2r/n} \left(\frac{4r}{n} \right)^{l}
L^{2l+1}_{n-l-1} \left(\frac{4r}{n} \right)
\end{displaymath}
where $L_{n}^{\alpha}(x)$ are associated Laguerre polynomials defined as
\begin{displaymath}
L_{n}^{\alpha}(x)=\frac{1}{n!} e^{x} x^{-\alpha} \frac{d^{n}}{dx^{n}}
\left( e^{-x} x^{n+\alpha} \right) =
\sum_{k=0}^{n} (-1)^{k}
\left(\begin{array}{c} n + \alpha\\ n-k \end{array} \right) \frac{x^{k}}{k!}.
\end{displaymath}
Note that this definition of the associated Laguerre polynomials while
consistent with standard mathematical usage~\cite{GR} differs from those
used in some standard physics texts~\cite{LF}.

In this basis the coefficients of the single particle wave functions are
defined via
\begin{eqnarray*}
c_{n,l,m,\tau}^{(1)}(t) & = & \langle n,l,m,\tau|\phi_{1}(t)\rangle \\
c_{n,l,m,\tau}^{(2)}(t) & = & \langle n,l,m,\tau|\phi_{2}(t)\rangle
\end{eqnarray*}
and the components of the full wavefunction $\Psi(t)$ in this basis will be
defined as
\begin{displaymath}
C_{n,l,m,\tau;n',l',m',\tau'}(t) = 
\beta M \left[ c_{n,l,m,\tau}^{(1)}(t) c_{n',l',m',\tau'}^{(2)}(t) -
c_{n',l',m',\tau'}^{(1)}(t) c_{n,l,m,\tau}^{(2)}(t) \right]
\end{displaymath}
which obviously incorporate the correct antisymmetry.

The basis~(\ref{basis}) is also used to expand the two-body interaction
$e^2/|{\bf r}_{1} - {\bf r}_{2}|$ in accord with~(\ref{2-body}).
Specifically, we calculated the matrix elements 
\begin{displaymath}
\langle i_{1};j_{1}| V(1,2) |i_{2};j_{2}\rangle =
\langle \psi_{n_{1},l_{1},m_{1}} (1) \psi_{n_{2},l_{2},m_{2}} (2)
|\frac{e^{2}}{|{\bf r}_{1}-{\bf r}_{2}|}|
\psi_{n_{1}',l_{1}',m_{1}'} (1) \psi_{n_{2}',l_{2}',m_{2}'} (2) \rangle
\end{displaymath}
for which we provide formulas in Appendix A. Here $i_{1}(n_{1},l_{1},m_{1})$,
$j_{1}(n_{2},l_{2},m_{2})$, $i_{2}(n_{1}',l_{1}',m_{1}')$ and
$j_{2}(n_{2}',l_{2}',m_{2}')$ are composite integer indices ranging from
one to infinity (or the maximum number of elements in the basis set).
The procedure outlined in section~\ref{dec} was then performed numerically
to obtain the expansion of the two-body interaction. If due to truncation of
the basis set $1\leq i_{1},j_{1},i_{2},j_{2} \leq K$, then the number of
terms in the decomposition~(\ref{2-body}) is $p=K^{2}$. In practice it is
convenient to consider some $n_{max}$ from which it follows that
$K=n_{max}(n_{max}+1)(2n_{max}+1)/6$.

The stochastic equations for the coefficients thus take the form
\begin{equation}
\begin{array}{l}
\displaystyle
dc_{n,l,m,\tau}^{(1)}(t) =
\left(\sum_{s=1}^{p} \langle O_{s}\rangle_{1} \left[ \frac{i}{2}
\langle O_{s} \rangle_{2} \omega_{s} dt
-\sqrt{-i\omega_{s}}dW_{s}(t)\right] + \frac{2i}{n^{2}} dt \right) \;
c_{n,l,m,\tau}^{(1)}(t) \\
\displaystyle
- \left( \sum_{s=1}^{p} \left[ i\langle O_{s}\rangle_{2} \omega_{s} dt
- \sqrt{-i\omega_{s}} dW_{s}(t)\right] \right)
\sum_{n',l',m'} \langle n,l,m|O_{s}|n',l',m'\rangle \;
c_{n',l',m',\tau}^{(1)}(t) \\
\displaystyle
- \left( \sum_{s=1}^{p}| \omega_{s}|
\frac{\langle \phi_{1}|\phi_{1} \rangle \left[ \langle O_{s}^{\dag}
O_{s}\rangle_{1} - |\langle O_{s}\rangle_{1}|^{2}\right]}
{2{\rm Re}\{\langle \phi_{1}|\phi_{2}\rangle\}} dt \right) \;
c_{n,l,m,\tau}^{(2)}(t)\\
\displaystyle
dc_{n,l,m,\tau}^{(2)}(t) =
\left(\sum_{s=1}^{p} \langle O_{s} \rangle_{2} \left[ \frac{i}{2}
\langle O_{s} \rangle_{1} \omega_{s} dt 
-\sqrt{-i\omega_{s}}dW_{s}(t) \right] + \frac{2i}{n^{2}} dt \right) \;
c_{n,l,m,\tau}^{(2)}(t) \\
\displaystyle
- \left( \sum_{s=1}^{p} \left[ i\langle O_{s} \rangle_{1} \omega_{s} dt
- \sqrt{-i\omega_{s}} dW_{s}(t)\right] \right)
\sum_{n',l',m'} \langle n,l,m|O_{s}|n',l',m'\rangle \;
c_{n',l',m',\tau}^{(2)}(t) \\
\displaystyle
- \left( \sum_{s=1}^{p}| \omega_{s}|
\frac{\langle \phi_{2}|\phi_{2} \rangle \left[ \langle O_{s}^{\dag}
O_{s}\rangle_{2} - |\langle O_{s}\rangle_{2}|^{2}\right]}
{2{\rm Re}\{\langle \phi_{1}|\phi_{2}\rangle\}} dt \right) \;
c_{n,l,m,\tau}^{(1)}(t)
\end{array}
\label{HeEqs}
\end{equation}
where we calculate expectations via formulas such as
\begin{displaymath}
\langle \phi_{1}|O_{s}|\phi_{1} \rangle
=\sum_{\tau} \sum_{n,l,m} \sum_{n',l',m'} c_{n,l,m,\tau}^{(1)*}(t)
\langle n,l,m|O_{s}|n',l',m'\rangle c_{n',l',m',\tau}^{(1)}(t).
\end{displaymath}

\begin{figure}[htp]
\begin{center}
\epsfig{file=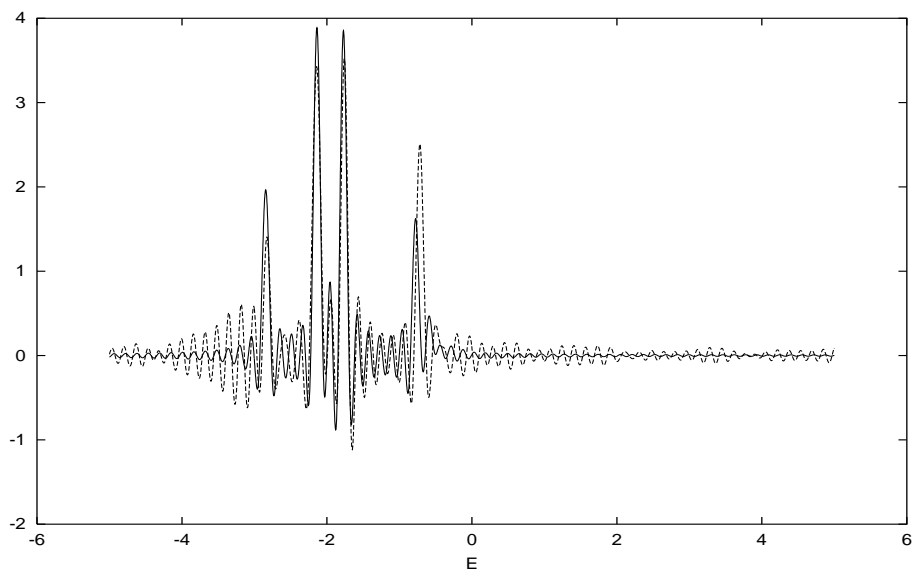,width=5.in,height=3.in}
\caption{He energy spectrum}
\label{hesp}
\end{center}
\end{figure}

An initial state consisting of random mixtures of 1s and 2s He$^{+}$ basis
functions for each electron was chosen. We chose a basis set with $n_{max}=4$
to perform the calculations. Equations~(\ref{HeEqs}) were solved using an
order 4.5 variable time-step (i.e., adaptive) Runge-Kutta method which has
been specifically developed to solve such stochastic differential
equations~\cite{Wilk2,Wilk3}. 
\begin{figure}[htp]
\begin{center}
\epsfig{file=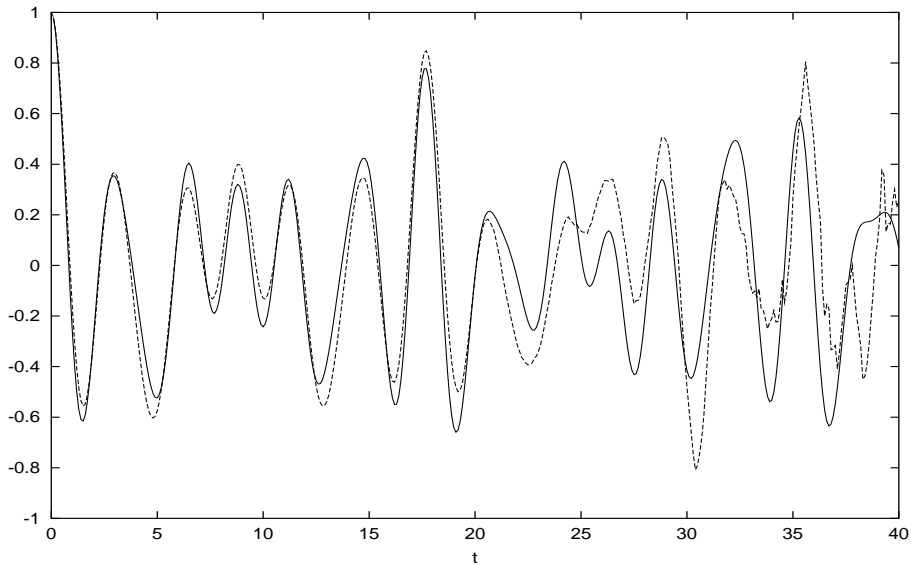,width=5.in,height=3.in}
\caption{Re $\langle \Psi(0)|\Psi(t)\rangle$ vs. $t$}
\label{repsi}
\end{center}
\end{figure}
\begin{figure}[htp]
\begin{center}
\epsfig{file=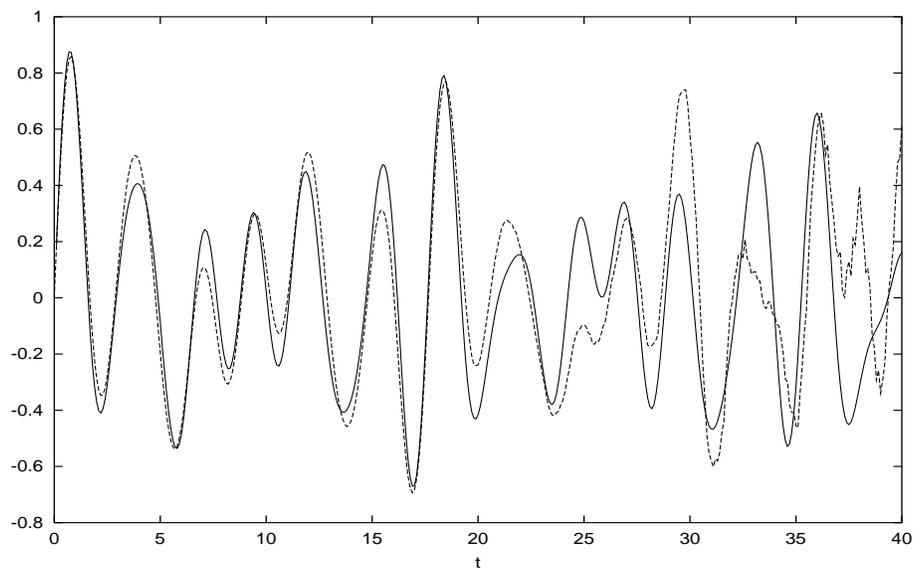,width=5.in,height=3.in}
\caption{Im $\langle \Psi(0)|\Psi(t)\rangle$ vs. $t$}
\label{impsi}
\end{center}
\end{figure}
A detailed discussion of the computational method will be presented
elsewhere~\cite{Wilk3}. In Figs.~\ref{repsi} and~\ref{impsi} we plot the
real and imaginary (respectively) parts of $\langle \psi(0)|\psi(t)\rangle$
against time in atomic units for an exact propagation of the initial state
(solid curve) and for the solution obtained via Eqs.~(\ref{HeEqs}) for
200000 realisations (dashed curve). The agreement is satisfactory although
the calculation has not completely converged.
In Fig.~\ref{hesp} we show the energy spectrum calculated via
Eq.~(\ref{spectrum}) for the exact and stochastic wave solutions.
Again agreement is good with the stochastic calculation reproducing all
energy levels..

\section{Extension to bosons}
\label{bos}

Stochastic decompositions for pairwise interactions are also of interest
for bosons in the context of Bose-Einstein condensation~\cite{CCD}. Here we
show that our approach can be adapted to bosons as well as fermions.

To begin with consider the case of two pairwise interacting bosons. Again we
break the {\em general} initial wavefunction into a sum of symmetric states
of the form $|\Psi(0)\rangle = \beta(|\phi_{1}(0)\rangle |\phi_{2}(0)\rangle
+|\phi_{2}(0)\rangle |\phi_{1}(0)\rangle )$. We need an exact means of
propagating the single particle states individually such that
\begin{equation}
|\Psi(t)\rangle = \beta M\left[|\phi_{1}(t)\rangle |\phi_{2}(t)\rangle
+ |\phi_{2}(t)\rangle |\phi_{1}(t)\rangle \right]
\label{true}
\end{equation}
where $|\phi_{1}(t)\rangle$ and $|\phi_{2}(t)\rangle$ satisfy norm-preserving
stochastic wave equations. To do this we add a fictitious subsystem of two
spin-1/2 degrees of freedom with null Hamiltonian and anti-symmetric state
$(1/\sqrt{2})(|+\rangle |-\rangle-|-\rangle|+\rangle)$ to our problem.
We thus have a total wavefunction
\begin{equation}
\begin{array}{ccl}
|\Psi_{fict}(t)\rangle & = & \displaystyle
(\beta/\sqrt{2}) M\left[ |\phi_{1}(t)\rangle |\phi_{2}(t)\rangle +
|\phi_{2}(t)\rangle |\phi_{1}(t)\rangle \right] \otimes
\left( |+\rangle |-\rangle-|-\rangle|+\rangle \right) \\
& = & \displaystyle
(\beta/\sqrt{2}) M \left[ \left( |\phi_{1+}(t)\rangle |\phi_{2-}(t)\rangle -
|\phi_{2-}(t)\rangle |\phi_{1+}(t)\rangle \right) \right. \\
& - & \displaystyle \left.
\left(|\phi_{1-}(t)\rangle |\phi_{2+}(t)\rangle -
|\phi_{2+}(t)\rangle |\phi_{1-}(t)\rangle \right) \right]
\end{array}
\label{fict}
\end{equation}
where $|\phi_{i\pm}(t)\rangle=|\phi_{i}\rangle |\pm\rangle$ for $i=1,2$.
This wavefunction is a sum of two antisymmetric states. It is thus clear
that solutions of~(\ref{fict}) can be obtained by determining the time
evolution of two-particle antisymmetric states
\begin{equation}
|\phi_{1\sigma_1}(t)\rangle |\phi_{2\sigma_2}(t)\rangle -
|\phi_{2\sigma_2}(t)\rangle |\phi_{1\sigma_1}(t)\rangle,
\label{fict2}
\end{equation}
for $\sigma_1,\sigma_2=\pm$, which can be obtained with the method for
fermions outlined above. Hence we can obtain~(\ref{fict}) at the cost of
including an extra $2$ component spin to each single particle state.
From~(\ref{fict}) we can get~(\ref{true}) by projecting out the fictitious
part of the solution via
\begin{displaymath}
|\Psi(t)\rangle = \frac{1}{\sqrt{2}}
(\langle +|\langle -| - \langle -|\langle +|)|\Psi_{fict}(t)\rangle.
\end{displaymath}
Hence the 2-boson problem can be solved using the 2-fermion formalism at
the expense of doubling the number of equations.

For an arbitrary number of bosons $N$ we wish to find a stochastic
decomposition
\begin{displaymath}
|\Psi(t)\rangle = \beta M \left[ S|\phi_{1}(t)\rangle |\phi_{2}(t)\rangle
\ldots |\phi_{N}(t)\rangle \right]
\end{displaymath}
for dynamics generated by Hamiltonian~(\ref{HAM}) where $S$ is the
symmetrisation operator. Let 
\begin{displaymath}
|a\rangle = \alpha A_{a}|\sigma_{1}\rangle \ldots |\sigma_{N}\rangle,
\end{displaymath}
where $|\sigma_{j}\rangle$ denotes one of a set of $N$ spin states, and
$A_{a}$ is the anti-symmetrisation operator on this space. Here $\alpha$ is
a normalization constant. If the spins again have a null Hamiltonian then
we may define a fictitious dynamics
\begin{displaymath}
|\Psi_{fict}(t)\rangle = |\Psi(t)\rangle |a\rangle =
\alpha \beta M \left[S A_{a} |\phi_{1\sigma_{1}}\rangle \ldots
|\phi_{N\sigma_{N}}\rangle \right]
\end{displaymath}
where each of the $N!$ terms
$A_{a} |\phi_{1\sigma_{1}}\rangle \ldots |\phi_{N\sigma_{N}}\rangle$
in the symmetrisation sum is antisymmetric.
Here $|\phi_{j\sigma_{j}}\rangle = |\phi_{j}\rangle |\sigma_{j}\rangle$ and
so by adding an extra fictitious spin with $N$ allowed states we can convert
the problem into fermion form. Application of the fermion method is then
straightforward and the boson wavefunction can be extracted in the end by
projecting out the fictitious spin state via
$|\Psi(t)\rangle = \langle a|\Psi_{fict}(t)\rangle$.
Because of the need to introduce a fictitious spin the computational costs
of the boson method scale between $O(N^{3})$ and $O(N^{6})$ depending on the
nature of the interaction.

\section{Summary}

We have shown that the time-dependent quantum $N$-body problem for pairwise
interacting fermions can be exactly decomposed into $N$ one-body problems
each of which obeys a stochastic norm-conserving wave equation. Our
approach improves on previous decompositions~\cite{JC} because the single
particle equations conserve norm and thus are much more stable numerically.
Use of the method was demonstated by calculating the low energy spectrum of
Helium. We have also explained how the approach can be extended to bosons.

The authors acknowledge the financial support of the Natural
Sciences and Engineering Research Council of Canada.

\section{Appendix A}

Consider the single particle Hamiltonian $H=-\hbar^{2} \nabla^{2}/2m_{e}
- Z e^{2}/r$ which has Hydrogen-like eigenfunctions of the form~(\ref{basis})
with
\begin{displaymath}
R_{n,l}(r) = \frac{2}{n^{2}} \sqrt{Z^{3} \frac{(n-l-1)!}{(n+l)!}}
e^{-Zr/n} \left(\frac{2Zr}{n}\right)^l
L^{2l+1}_{n-l-1}\left(\frac{2Zr}{n}\right)
\end{displaymath}
in atomic units with associated energies $E_{n} = -Z^{2}/2n^{2}$.
It can then be shown that
\begin{displaymath}
\begin{array}{cl}
& \displaystyle
\langle \psi_{n_{1},l_{1},m_{1}} (1) \psi_{n_{2},l_{2},m_{2}} (2)
|\frac{e^{2}}{|{\bf r}_{1}-{\bf r}_{2}|}|
\psi_{n_{1}',l_{1}',m_{1}'} (1)\psi_{n_{2}',l_{2}',m_{2}'} (2)\rangle \\
= & \displaystyle
\frac{Z}{16} \sqrt{\frac{(2l_{1}'+1)(2l_{2}'+1)}{(2l_{1}+1)(2l_{2}+1)}}
\sqrt{\frac{(n_{1}-l_{1}-1)!(n_{1}'-l_{1}'-1)!(n_{2}-l_{2}-1)!
(n_{2}'-l_{2}'-1)!}
{(n_{1}+l_{1})!(n_{1}'+l_{1}')!(n_{2}+l_{2})!(n_{2}'+l_{2}')!}} \\
\times & \displaystyle
\sum_{l=0}^{\infty} \sum_{m=-l}^{l} (-1)^{m}
\left[ \begin{array}{ccc} l_{1}' & l & l_{1} \\
                          m_{1}' & m & m_{1}
     \end{array} \right]
\left[ \begin{array}{ccc} l_{2}' & l & l_{2} \\
                          m_{2}' &-m & m_{2}
       \end{array}\right]
\left[ \begin{array}{ccc} l_{1}' & l & l_{1} \\
                          0      & 0 & 0
       \end{array}\right]
\left[ \begin{array}{ccc} l_{2}' & l & l_{2} \\
                          0      & 0 & 0
       \end{array}\right] \\
\times & \displaystyle
\sum_{k_{1}=0}^{n_{1}-l_{1}-1} \sum_{k_{1}'= 0}^{n_{1}'- l_{1}'-1}
\sum_{k_{2}=0}^{n_{2}-l_{2}-1} \sum_{k_{2}'= 0}^{n_{2}'- l_{2}'-1}
(-1)^{k_{1} + k_{1}' + k_{2} + k_{2}'}
\frac{(l_{1}+l_{1}'+l_{2}+l_{2}'+k_{1}+k_{1}'+k_{2}+k_{2}'+4)!}
{k_{1}!k_{1}'!k_{2}!k_{2}'!} \\
\times & \displaystyle
\left(\begin{array}{c} n_{1}+l_{1} \\
                       n_{1}-l_{1}-1-k_{1}
      \end{array} \right)
\left(\begin{array}{c} n_{1}'+l_{1}' \\
                       n_{1}'-l_{1}'-1-k_{1}'
      \end{array} \right)
\left(\begin{array}{c} n_{2}+l_{2} \\
                       n_{2}-l_{2}-1-k_{2}
      \end{array} \right)
\left(\begin{array}{c} n_{2}'+l_{2}' \\
                       n_{2}'-l_{2}'-1-k_{2}'
      \end{array} \right) \\
\times &  \displaystyle
\frac{(2/n_{1})^{k_{1}+l_{1}+2}(2/n_{1}')^{k_{1}'+l_{1}'+2}
(2/n_2)^{k_{2}+l_{2}+2}(2/n_2')^{k_{2}'+l_{2}'+2}}
{\left(1/n_{1}+1/n_{1}'+1/n_{2}+1/n_{2}'\right)^{l_{1}+l_{1}'+l_{2}+l_{2}'+
k_{1}+k_{1}'+k_{2}+k_{2}'+5}} \\
\times & \displaystyle
\left[ \frac{1}{l+l_{1}+l_{1}'+k_{1}+k_{1}'+3}
F(1,l_{1}+l_{1}'+l_{2}+l_{2}'+k_{1}+k_{1}'+k_{2}+k_{2}'+5; \right. \\
& \displaystyle
l+l_{1}+l_{1}'+k_{1}+k_{1}'+4;
\frac{1/n_{1}+1/n_{1}'}{1/n_{1}+1/n_{1}'+1/n_{2}+1/n_{2}'}) \\
+ & \displaystyle
\frac{1}{l+l_{2}+l_{2}'+k_{2}+k_{2}'+3}
F(1,l_{1}+l_{1}'+l_{2}+l_{2}'+k_{1}+k_{1}'+k_{2}+k_{2}'+5; \\
 & \displaystyle
\left. l+l_{2}+l_{2}'+k_{2}+k_{2}'+4;
\frac{1/n_{2}+1/n_{2}'}{1/n_{1}+1/n_{1}'+1/n_{2}+1/n_{2}'}) \right]
\end{array}
\end{displaymath}
where $F(a,b;c;d)$ is the hypergeometric function~\cite{GR}. Here 
\begin{displaymath}
\left[\begin{array}{ccc} j_{1} & j_{2} & j \\
                         m_{1} & m_{2} & m
      \end{array}\right]
= \langle j_{1},j_{2},m_{1},m_{2}|j,m \rangle
\end{displaymath}
denote the Clebsch-Gordon coefficients. We have used the convention of
Ref.~\cite{Dev} in which
\begin{equation}
\left[\begin{array}{ccc} j_{1} & j_{2} & j \\
                         m_{1} & m_{2} & m
      \end{array}\right]
=\delta_{m,m_{1}+m_{2}} \sqrt{(2j+1)AB} \sum_{n=0}^{\infty}
\frac{(-1)^{n}}{n!C_{n}}
\label{sumf}
\end{equation}
with
\begin{displaymath}
\begin{array}{ccl}
A & = & \displaystyle
\frac{(j_{1}+j_{2}-j)! (j+j_{1}-j_{2})! (j+j_{2}-j_{1})!}
{(j+j_{1}+j_{2}+1)!} \\
B & = & \displaystyle
(j_{1}+m_{1})! (j_{1}-m_{1})! (j_{2}+m_{2})! (j_{2}-m_{2})!(j+m)!(j-m)! \\
C_{n} & = & \displaystyle
(j_{1}+j_{2}-j-n)! (j_{1}-m_{1}-n)! (j_{2}+m_{2}-n)!(j-j_{2}+m_{1}+n)! \\
& \times & \displaystyle
(j-j_{1}-m_{2}+n)! 
\end{array}
\end{displaymath}
where it is understood that the sum in~(\ref{sumf}) truncates when factorials
have negative arguments.

\end{document}